\documentclass[prl,letterpaper,aps,floatfix,twocolumn]{revtex4}
\usepackage{graphicx}
\usepackage{amsmath}
\newcommand{\beq}{\begin{equation}}
\newcommand{\eeq}{\end{equation}}
\newcommand{\bk}{{{\bf{k}}}}

\newcommand{\br}{{{\bf{r}}}}

\newcommand{\bq}{{\bf{q}}}

\newcommand{\beqa}{\begin{eqnarray}}
\newcommand{\eeqa}{\end{eqnarray}}

\newcommand{\pdg}{{\vphantom\dag}}
\begin{document}
\title{Spin and Charge Transport on the Surface of a Topological Insulator}
\author{A.A. Burkov and D.G. Hawthorn}
\affiliation{Department of Physics and Astronomy, University of Waterloo, Waterloo, Ontario 
N2L 3G1, Canada}
\date{\today}
\begin{abstract}
We derive diffusion equations, which describe spin-charge coupled transport on 
the helical metal surface of a three-dimensional topological insulator. 
The main feature of these equations is a large magnitude of the spin-charge 
coupling, which leads to interesting and observable effects. In particular, we predict a new magnetoresistance effect, which manifests 
in a nonohmic correction to a voltage drop between a ferromagnetic spin-polarized electrode 
and a nonmagnetic electrode, placed on top of the helical metal. 
This correction is proportional to the cross-product 
of the spin polarization of the ferromagnetic electrode and the charge current between 
the two electrodes. We also demonstrate tunability of this effect by applying a gate voltage, 
which makes it possible to operate the proposed device as a transistor. 
\end{abstract}
\maketitle
Time-reversal invariant topological insulator (TI) is a new state of matter, distinguished from a regular 
band insulator by a nontrivial topological invariant, which characterizes 
its bandstructure \cite{Kane10}.
Its theoretical \cite{Kane05} and experimental 
\cite{Konig07} discovery has accordingly generated a great deal of excitement in the 
condensed matter physics community. 
The most robust observable consequence of a nontrivial topological character of these materials
is the presence of gapless helical edge states, whose gaplessness is protected by
time-reversal symmetry and is thus robust to perturbations that do not break this 
symmetry. 
In particular, the surface of a three-dimensional (3D) TI, such as 
$\textrm{Bi}_2\textrm{Se}_3$ or $\textrm{Bi}_2\textrm{Te}_3$ \cite{Hasan09}, is a 2D metal, whose bandstructure consists of 
an odd number of Dirac cones, centered at time-reversal invariant momenta
in the surface Brillouin zone  \cite{Kane05}.
Assuming the Fermi surface encloses only one Dirac point, the low-energy Hamiltonian, describing
such a 2D metal, is given by:
\beq
\label{eq:1}
H = \sum_{\bk} v_F (\bk \times \hat z) \cdot {\boldsymbol \tau}_{\sigma \sigma'}
c^{\dag}_{\bk \sigma} c^{\pdg}_{\bk \sigma'}, 
\eeq
where $\hat z$ is along the normal to the surface, $v_F$ is the Fermi velocity, ${\boldsymbol \tau}$ is the vector of Pauli matrices, summation over repeated spin indices is implicit, and we use $\hbar =1$ units henceforth. 
The eigenstates of this Hamiltonian are labeled by {\em helicity}, i.e. projection of the 
spin of the electron on the direction of its momentum, hence the name {\em helical metal} (HM).  
The most obvious physical property of a HM is a strong coupling between the spin and orbital degrees of freedom, the energy scale characterizing this coupling being the Fermi 
energy $\epsilon_F$. Spin-orbit (SO) coupling of such a magnitude is unprecedented 
among known materials and it is thus extremely interesting to work out its possible 
observable consequences. Some work on this subject has already appeared in the literature ~\cite{Raghu10,Franz09}.

In this Letter we will focus on spin and charge transport phenomena in a 2D HM. 
In this setting, it is useful to note that the Hamiltonian of a HM Eq.(\ref{eq:1}) is 
very similar to the Rashba Hamiltonian \cite{Rashba60}, which has been studied 
extensively in the context of spin transport in semiconductor-based two-dimensional electron gas (2DEG) 
systems \cite{Perel71,Burkov04,Weber07}. 
The main difference is the magnitude of the SO term: while in a semiconductor 2DEG it is only a weak perturbation 
on top of the band kinetic energy, it is the only term present in the Hamiltonian of the HM.  
Thus, while the spin-charge coupling effects in semiconductor 2DEG's are extremely weak and  
largely experimentally unobservable, they can be expected to be significant in a HM. 

We start from the HM Hamiltonian Eq.(\ref{eq:1}). We will assume that the surface Fermi energy, measured relative to the 
Dirac point, is finite (we will take it to be positive for concreteness), as is generically the case in real materials \cite{Kane10}, and that the corresponding 2D Fermi surface encloses one Dirac point (we expect our results to hold for any odd number of Dirac cones as 
different cones will contribute additively to transport).
We will assume that nonmagnetic impurities with potential $V_i(\br) = u_0 \sum_a \delta(\br - \br_a)$ and 2D density $n_i$ are present at the surface. 
To derive the diffusive transport equation we will adopt the formalism of Ref.~\cite{Burkov04}, based on the evaluation of the 
density matrix response function. 

The real-time disorder-averaged Green's function of a 2D HM is a $2 \times 2$ matrix in spin space.
It is convenient to write it as a sum of ``singlet" and ``triplet" contributions $G^{R,A}_{\sigma \sigma'}(\bk, \omega) = G^{R,A}_s(\bk, \omega) \delta_{\sigma \sigma'} + {\bf  G}^{R,A}_t(\bk, \omega) 
\cdot {\boldsymbol \tau}_{\sigma \sigma'}$,
where $R,A$ stand for {\em retarded} and {\em advanced} and the singlet and triplet Green's functions, projected onto 
the band with positive helicity, are given by:
\beqa
\label{eq:3}
G^{R,A}_s(\bk, \omega)&=&\frac{1/2}{\omega - \xi_\bk \pm \frac{i}{2 \tau}}, \nonumber \\
{\bf G}^{R,A}_t(\bk, \omega)&=&\frac{\hat k \times \hat z}{2} \frac{1}{\omega - \xi_\bk \pm \frac{i}{2 \tau}}. 
\eeqa
Here $\xi_\bk = v_F |\bk| - \epsilon_F$ and the impurity scattering rate is given by 
$1/\tau = \pi n_i u_0^2 \rho(\epsilon_F)$,
where $\rho(\epsilon_F) = \epsilon_F/ 2 \pi v_F^2$ is the density of states at Fermi energy. 
We assume that $\epsilon_F \tau \gg 1$, which allows us to treat disorder perturbatively.  
We then introduce generalized density operators $\varrho_{\sigma \sigma'}(\br, t) = \Psi^\dag_{\sigma'}(\br, t) \Psi^\pdg_{\sigma}(\br, t)$, 
and evaluate the retarded density response function $\chi_{\sigma_1 \sigma_2, \sigma_3 \sigma_4}(\br - \br', t - t') = - i \theta(t - t') 
\langle [\varrho^\pdg_{\sigma_1 \sigma_2}(\br, t), \varrho^\dag_{\sigma_3 \sigma_4}(\br', t')] \rangle$.
To leading order in the small parameter $1/\epsilon_F \tau$ this is easily evaluated by summing all ladder vertex 
corrections to the polarization bubble diagram \cite{Burkov04}.
Assuming $q \ll k_F$ and $\Omega \ll \epsilon_F$, one obtains the following result for the 
Fourier transformed density response function: $\chi(\bq,\Omega)= \chi^0 - i\Omega \tau \rho(\epsilon_F)
{\cal I}(\bq,\Omega) {\cal D}(\bq,\Omega)$,
where $\chi^0$ is the static uniform susceptibility and  ${\cal I}_{\sigma_1\sigma_2,\sigma_3\sigma_4}(\bq,\Omega)= 
n_i u_0^2 \int \frac{d^2 k}{(2\pi)^2} G^R_{\sigma_1\sigma_3}(\bk + \bq, \Omega)
G^A_{\sigma_4\sigma_2}(\bk,0)$.
${\cal D} = (1-{\cal I})^{-1}$ is the diffusion propagator, i.e. the Green's function of the spin-charge coupled transport equation we seek to derive.   
Evaluating the momentum integral above, expanding the result to leading nontrivial order in $i \Omega \tau$ and $v_F q \tau$, which corresponds physically to coarse-graining over 
length scale of the order of the mean free path $\ell = v_F \tau$ and time scale of order $\tau$, 
we finally obtain the following set of spin-charge coupled transport equations:
\begin{eqnarray}
\label{eq:9}
\frac{\partial N}{\partial t}&=& D{\boldsymbol \nabla}^2  N
+ 2 \Gamma  (\hat{z} \times {\boldsymbol \nabla})
 \cdot {\bf S}, \nonumber \\
\frac{\partial S^x}{\partial t}&=&\frac{D}{2}\frac{\partial^2 S^x}{\partial x^2} + 
\frac{3 D}{2} \frac{\partial^2 S^x}{\partial y^2} + D \frac{\partial^2 S^y}
{\partial x \partial y} \nonumber \\
&-&\frac{S^x}{\tau} + \Gamma (\hat z \times {\boldsymbol \nabla})_x N, \nonumber \\
\frac{\partial S^y}{\partial t}&=&\frac{D}{2}\frac{\partial^2 S^y}{\partial y^2} + 
\frac{3 D}{2} \frac{\partial^2 S^y}{\partial x^2} + D \frac{\partial^2 S^x}
{\partial x \partial y} \nonumber \\
&-&\frac{S^y}{\tau} + \Gamma (\hat z \times {\boldsymbol \nabla})_y N. 
\end{eqnarray}       
Here ${\bf S} $ is the nonequilibrium spin density and $N/\rho(\epsilon_F)$ is the full local electrochemical 
potential (we have subsumed the contribution of the external electrostatic potential into the definition of the charge density $N$).  
$D = v_F^2 \tau/2$ is the diffusion constant and $\Gamma = \sqrt{D / 2 \tau} = v_F/2 $ is the 
spin-charge coupling constant. 
The out-of-plane component of the spin density does not appear in the above equations.
The reason for this is that any nonequilibrium spin polarization in the $z$-direction 
will precess with frequency of order $\epsilon_F$ around the momentum-dependent 
SO field $\hat z \times \bk$ and will thus average out to zero on time scales longer than
$1/\epsilon_F$. 

It is useful to compare Eq.(\ref{eq:9}) with the corresponding equations for semiconductor 
2DEG systems with Rashba SO interactions, derived in Ref.\cite{Burkov04}. 
In the latter case, the coupling between the in-plane and out-of-plane spin components 
is large, of order $\lambda \epsilon_F \tau$, where $\lambda$ is the Rashba SO coupling constant.
This leads to interesting and experimentally observable 
effects in spin transport \cite{Weber07}. These effects will not be observable in HM due to a vanishing coupling between in-plane and out-of-plane spin components. 
In contrast, the spin-charge coupling in Rashba 2DEG systems is extremely small, 
with $\Gamma \sim \lambda (\lambda k_F \tau)^2$, and is largely experimentally unobservable. 
In a HM, on the other hand, the spin-charge coupling is much stronger: 
$\Gamma \sim v_F$, and one can thus expect these effects to be easily experimentally accessible. 
Another difference from the Rashba 2DEG case is an anisotropic nature of spin diffusion
in the HM: the diffusion constant is different in the direction of the spin polarization 
and perpendicular to it. This is not unexpected and is also a consequence of the strong SO 
coupling in this system. 
Finally, the relaxation time for the in-plane spin components in the HM case is 
the same as the momentum relaxation time $\tau$.
In the Rashba 2DEG case, in contrast, the relaxation 
time is much longer, of order $\tau/(\lambda k_F \tau)^2$. 
\begin{figure}
\includegraphics[width=7cm]{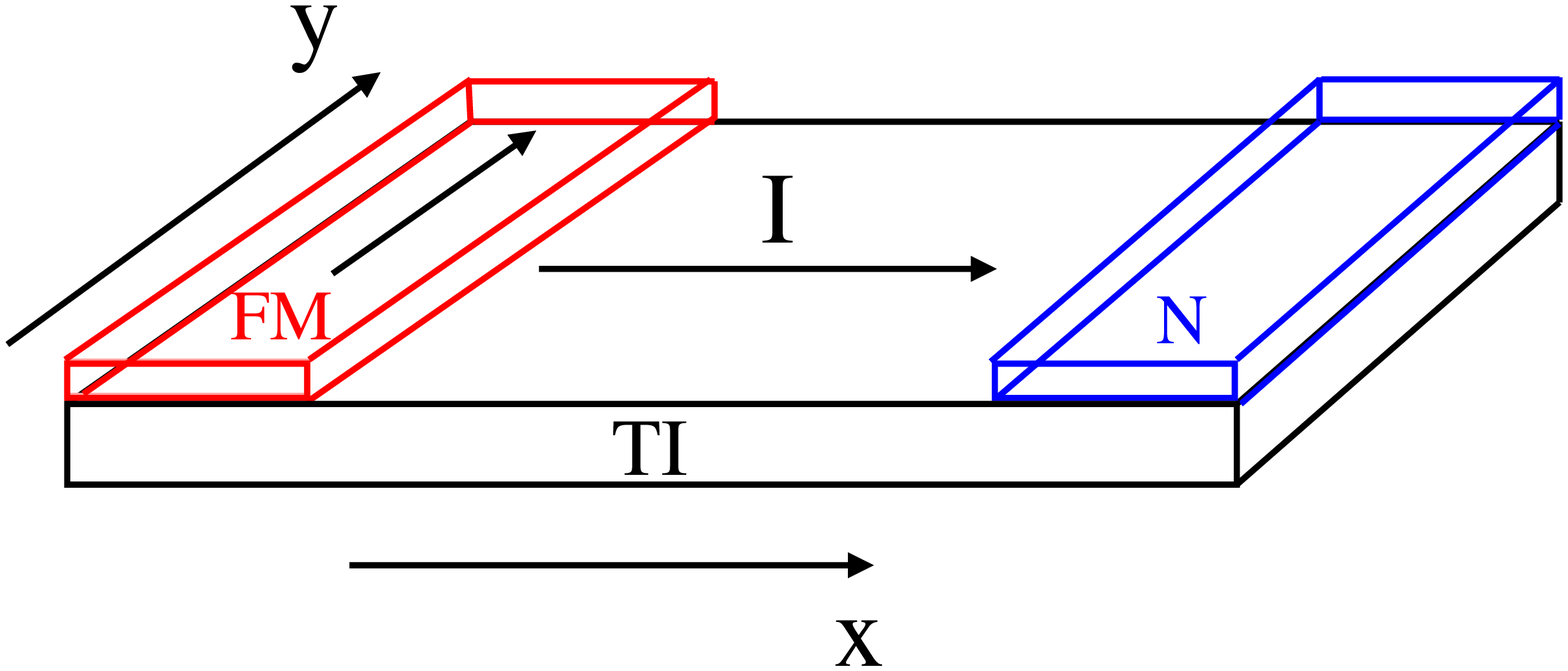}
\caption{(Color online). Schematic picture of the sample. The left electrode (FM) is ferromagnetic and can inject 
a spin-polarized current (polarization shown by arrow) into the sample. The right electrode (N) is nonmagnetic. The voltage 
drop between the electrodes has a contribution, proportional to the cross-product of the
polarization and the current.}
\end{figure}

Unusual spin-charge transport properties of the HM are embodied
in a nonstandard relation between the charge current and the gradient of the 
electrochemical potential, which are no longer simply proportional to each other.  
This can be seen from the first of Eq.(\ref{eq:9}). 
Namely, since the electron charge is conserved, the charge transport equation 
must have the form:
\beq
\label{eq:10}
\frac{\partial N}{\partial t} = - {\boldsymbol \nabla} \cdot {\bf J}, 
\eeq
where ${\bf J}$ is the particle current density (equal to $1/e \times$ the electrical current 
density). Then from Eq.(\ref{eq:9}) we can immediately read off the following 
expression for the particle current density:
\beq
\label{eq:11}
{\bf J} = - D {\boldsymbol \nabla} N + 2 \Gamma (S^x \hat y - S^y \hat x). 
\eeq
The first term here is the usual contribution to the charge current, proportional to the 
charge electrochemical potential gradient. 
There is, however, a second term, which is proportional to the excess spin electrochemical 
potentials, if the spins are polarized in-plane (in the case of a clean TI surface this term was 
first studied in Ref.\cite{Raghu10}). The physical origin of this term is a definite helicity of the Fermi surface states 
in a HM.

We will now show that this extra term in the expression for the charge current manifests as a novel
and potentially useful magnetoresistance effect.  
We imagine a setup, schematically depicted in Fig.1. Assume that two 
electrodes are deposited on top of a TI sample a distance $L$ from each other. 
Assume the left electrode is ferromagnetic (FM) and can inject a spin-polarized current into the 
surface of the sample, while the right electrode is nonmagnetic (N).
Let a fixed current $I$ flow between 
the electrodes. We will now use Eq.(\ref{eq:9}) to calculate the corresponding voltage drop 
between the electrodes as a function of the degree and direction of the spin polarization 
in the FM electrode. 
Assuming there is no variation in the spin and charge electrochemical potentials in the 
$y$-direction (i.e. parallel to the electrodes), the spin-charge diffusion equations simplify to:
\beqa
\label{eq:12}
&&D \frac{d^2  N}{d x^2} + 2 \Gamma \frac{d S^y}{d x} = 0, \nonumber \\
&&\frac{3 D}{2} \frac{d^2  S^y}{d x^2} - \frac{S^y}{\tau} + \Gamma 
\frac{d  N}{d x} = 0. 
\eeqa  
We will assume for simplicity that the distance between the electrodes $L$ is greater than the spin diffusion length in the HM, 
which is of order $\ell$. Then we can assume that the spin transmission from the FM to 
the N electrode is zero and take the boundary conditions for Eq.(\ref{eq:12}) to be: 
\beqa
\label{eq:13}
J |_{x = \pm L/2}&=&\frac{I}{e}, \nonumber \\
-\left. \frac{3 D}{2} \frac{d S^y}{dx} \right |_{x = -L/2}&=&\frac{I \eta}{e}, \,\,\, \left. \frac{d S^y}{dx} \right |_{x = L/2} = 0,
\eeqa 
where $J = - D d N /d x -2 \Gamma  S^y$ is the particle current in the $x$-direction and 
$-(3 D/2)  d S^y/dx$ is the spin current. Note that the term $-\Gamma N$, which, as may naively be concluded from 
examining Eq.(\ref{eq:12}), contributes to the 
spin current, should not in fact be 
included in the definition of the spin current.  This term is not zero even in equilibrium and 
thus represents an equilibrium spin current \cite{Rashba03}, which has no relation to measurable 
transport spin current. One arrives at the same conclusion from a formal derivation of the spin current boundary 
conditions~\cite{Malshukov05}.
The second boundary condition assumes that the current, injected into the HM from the FM electrode, is spin polarized with the degree of spin polarization parametrized by $\eta$. 
Note that $\eta$ can be either positive or negative, corresponding to spin polarization of the FM electrode along 
$y$ or $-y$ direction. 
At the N electrode, the spin current is taken to be zero, as discussed above. 
Note that the polarization dependence disappears from Eqs.(\ref{eq:12}), (\ref{eq:13}) if the 
FM electrode is polarized in the $x$-direction, i.e. along the direction of the current. 

We now solve Eqs.(\ref{eq:12}) with the boundary conditions~(\ref{eq:13}). 
The first of Eqs.(\ref{eq:12}) simply gives:
\beq
\label{eq:14}
D \frac{d N}{dx} + 2 \Gamma  S^y = const = -I/e. 
\eeq
Expressing $dN/dx$ in terms of $S^y$ and substituting into the second Eq.(\ref{eq:12}) we obtain:
\beq
\label{eq:15}
\frac{d^2 S^y}{d x^2} - \frac{8}{3 \ell^2} S^y = \frac{4 I}{3e v_F \ell^2}, 
\eeq
which needs to be solved with boundary conditions (\ref{eq:13}).
The solution is given by: 
\beq
\label{eq:16}
S^y(x) = \frac{I \eta}{e v_F} \sqrt{\frac{2}{3}}  \frac{\cosh \left(\frac{2 x - L}{\sqrt{3/2} \ell}\right)}{\sinh \left(\frac{2 L}{\sqrt{3/2} \ell}\right)} - 
\frac{I}{2e v_F}.
\eeq 
The first term in Eq.(\ref{eq:16}) corresponds to the spin density, injected from the FM electrode. 
This term decays exponentially away from the electrode on length scale of the order of the mean free path $\ell$. 
The second term, on the other hand, is constant and is due to the {\em magnetoelectric effect}: generation of nonequilibrium 
in-plane spin polarization by charge current. Note that the magnetoelectric coefficient does not depend on disorder and in this 
sense is universal. 
 
The voltage drop between the electrodes can now be calculated as:
\beq
\label{eq:17}
V = - \frac{1}{e \rho(\epsilon_F)} \int_{-L/2}^{L/2}\frac{d N }{dx} dx = \frac{2 \pi I L}{e^2 k_F \ell} + 
\frac{4 \pi I \eta}{e^2 k_F}.
\eeq 
Thus the voltage drop consists of two contributions. The first contribution corresponds to the usual Ohmic resistance, which
is proportional to the separation between the electrodes $L$ and inversely proportional to the conductivity. 
The second contribution depends on the degree {\em and direction} of the polarization of the FM electrode and constitutes 
a new magnetoresistance effect. Note that this contribution is proportional to the {\em cross-product} of the polarization of the 
FM electrode and the charge current (recall that it disappears when the spin polarization of the FM electrode is along the $\pm x$-directions). 
This distinguishes it from all other known magnetoresistance effects.   
The effect has some degree of universality, in the sense that it is independent of disorder and the separation between the electrodes 
and only depends 
on the spin polarization of the injected current and the Fermi momentum, characterizing the bandstructure of the HM.  
However, the degree of the injected spin polarization of course does depend on nonuniversal properties of the interface between the 
FM electrode and the HM.

It is interesting to note that when $L <  2 \ell$, the resistance, according to Eq.(\ref{eq:17}), may formally become negative when $\eta$ is negative. This may mean two things. 
One possibility is that this simply indicates that we are going outside of the regime of validity of the gradient expansion of the diffusion propagator or of the boundary conditions Eq.(\ref{eq:13}) at short length scales. 
Another, more interesting possibility is that this signifies a real physical phenomenon: a nonequilibrium transition into a zero-resistance state, reminiscent of the microwave-induced zero resistance states in 2DEG \cite{Mani02}. 
We leave the resolution of this issue to future study.

 It is important to realize that while Eq.(\ref{eq:17}) was obtained using a specific form of the spin current boundary conditions Eq.(\ref{eq:13}), our result is in fact largely independent of the boundary conditions. If instead of Eq.(\ref{eq:13}) we used a more general form 
for the boundary conditions, involving both spin densities and spin density gradients (which physically would correspond to significant
spin-flip scattering at the electrode-HM interfaces), only numerical coefficient of the magnetoresistance term in Eq.(\ref{eq:17}) would 
change, while its general form would remain the same.   

The magnetoresistance effect we propose is fundamentally different from other known 
magnetoresistance effects and is easily distinguishable from them.
The extra voltage drop, first found by Johnson and Silsbee \cite{Silsbee85}, which always  occurs at interfaces between ferromagnetic and paramagnetic metals and which results from differences in conductivities of the majority and minority spin electrons in ferromagnets, is an even function of the 
spin polarization of the ferromagnet (also true when the paramagnetic metal is helical), while it is an odd function for the effect we propose.  
The famous giant magnetoresistance effect \cite{Fert93} requires both electrodes  
to be FM and be separated by a distance, smaller than the spin diffusion length. 
The effect we describe here requires only one electrode to be FM and does not depend on the distance between the electrodes.

Yet another interesting feature of Eq.(\ref{eq:17}) is that the magnitude of the polarization-dependent 
term may be varied independently of the Ohmic term by applying an external gate voltage to the HM to tune the Fermi energy.
This is based on the observation that the product of the Fermi momentum and the mean free path
is in fact independent of the Fermi energy $k_F \ell = 2 v_F^2 / n_i u_0^2$, which makes 
the Ohmic term insensitive to changing the Fermi energy \cite{footnote}. 
The magnitude of the polarization-dependent term, however, does explicitly depend on the Fermi energy  as $\sim 1/\epsilon_F$.
The possibility of realizing this transistor-like effect 
depends strongly on how insulating the bulk TI material actually is. 
It may thus be difficult to realize with currently available 3D TI, in which the bulk 
is not really insulating \cite{Kane10}. We expect, however, that these problems will be 
resolved in the next generation of TI materials. 

In conclusion, we have proposed a new magnetoresistance effect that should occur on the HM surface of a 3D TI. 
The effect manifests itself in a nonohmic correction to the voltage drop between two electrodes, placed on top of the HM: 
a polarized FM electrode and a nonmagnetic electrode. The correction is proportional to the cross-product of the charge current 
between the two electrodes and the spin polarization of the FM electrode and thus changes sign when the direction 
of the spin polarization is reversed.  
We have also proposed a transistor-like device, based on 
this new magnetoresistance effect. 

\begin{acknowledgments}
We acknowledge useful discussions with Leon Balents, Arun Paramekanti and Enrico Rossi.  
Financial support from the NSERC of Canada and a University of Waterloo start-up grant is gratefully acknowledged.  
\end{acknowledgments}

\end{document}